\documentclass[12pt,english]{article}
\usepackage[T1]{fontenc}
\usepackage[latin9]{inputenc}
\usepackage{amsthm}
\usepackage{amsmath}
\usepackage{amssymb}
\usepackage{graphicx}

\makeatletter

\providecommand{\tabularnewline}{\\}

\numberwithin{equation}{section}
\numberwithin{figure}{section}
\theoremstyle{plain}
\newtheorem{thm}{\protect\theoremname}
  \theoremstyle{definition}
  \newtheorem{defn}{\protect\definitionname}

\makeatother

\usepackage{babel}
  \providecommand{\definitionname}{Definition}
\providecommand{\theoremname}{Theorem}

\begin{document}

\title{Deception by Design: Evidence-Based Signaling Games for Network Defense}

\author{Jeffrey Pawlick and Quanyan Zhu }
\maketitle
\begin{abstract}
Deception plays a critical role in the financial industry, online
markets, national defense, and countless other areas. Understanding
and harnessing deception - especially in cyberspace - is both crucial
and difficult. Recent work in this area has used game theory to study
the roles of incentives and rational behavior. Building upon this
work, we employ a game-theoretic model for the purpose of mechanism
design. Specifically, we study a defensive use of deception: implementation
of honeypots for network defense. How does the design problem change
when an adversary develops the ability to detect honeypots? We analyze
two models: cheap-talk games and an augmented version of those games
that we call cheap-talk games with evidence, in which the receiver
can detect deception with some probability. Our first contribution
is this new model for deceptive interactions. We show that the model
includes traditional signaling games and complete information games
as special cases. We also demonstrate numerically that deception detection
sometimes eliminate pure-strategy equilibria. Finally, we present
the surprising result that the utility of a deceptive defender can
sometimes increase when an adversary develops the ability to detect
deception. These results apply concretely to network defense. They
are also general enough for the large and critical body of strategic
interactions that involve deception. 


\emph{Author affiliations:} Polytechnic School of Engineering of New
York University%
\footnote{Author Contact Information: Jeffrey Pawlick (jpawlick@nyu.edu), Quanyan
Zhu (quanyan.zhu@nyu.edu), Department of Electrical and Computer Engineering,
Polytechnic School of Engineering of NYU, 5 MetroTech Center 200A,
Brooklyn, NY 11201%
} (NYU)

This work is in part supported by an NSF IGERT grant through the Center
for Interdisciplinary Studies in Security and Privacy (CRISSP) at
NYU.
\end{abstract}

\section{Introduction\label{sec:Introduction}}

Deception has always garnered attention in popular culture, from the
deception that planted a seed of anguish in Shakespeare's Macbeth
to the deception that drew viewers to the more contemporary television
series \emph{Lie to Me}. Our human experience seems to be permeated
by deception, which may even be engrained into human beings via evolutionary
factors \cite{key-30,key-29}. Yet humans are famously bad at detecting
deception \cite{key-14,key-26}. An impressive body of research aims
to improve these rates, especially in interpersonal situations. Many
investigations involve leading subjects to experience an event or
recall a piece of information and then asking them to lie about it
\cite{key-31,key-14,key-15}. Researchers have shown that some techniques
can aid in detecting lies - such as asking a suspect to recall events
in reverse order \cite{key-14}, asking her to maintain eye contact
\cite{key-15}, asking unexpected questions or strategically using
evidence \cite{key-32}. Clearly, detecting interpersonal deception
is still an active area of research.

While understanding interpersonal deception is difficult, studying
deception in cyberspace has its set of unique challenges. In cyberspace,
information can lack permanence, typical cues to deception found in
physical space can be missing, and it can be difficult to impute responsibility
\cite{key-18}. Consider, for example, the problem of identifying
deceptive opinion spam in online markets. Deceptive opinion spam consists
of comments made about products or services by actors posing as customers,
when they are actually representing the interests of the company concerned
or its competitors. The research challenge is to separate comments
made by genuine customers from those made by self-interested actors
posing as customers. This is difficult for humans to do unaided; two
out of three human judges in \cite{key-20} failed to perform significantly
better than chance. To solve this problem, the authors of \cite{key-20}
make use of approaches including a tool called the \emph{Linguistic
Inquiry Word Count}, an approach based on the frequency distribution
of part-of-speech tags, and third approach which uses a classification
based on \emph{n}-grams. This highlights the importance of an interdisciplinary
approach to studying deception, especially in cyberspace.

Although an interdisciplinary approach to studying deception offers
important insights, the challenge remains of putting it to work in
a quantitative framework. In behavioral deception experiments, for
instance, the incentives to lie are also often poorly controlled,
in the sense that subjects may simply be instructed to lie or to tell
the truth \cite{key-19}. This prohibits a natural setting in which
subjects could make free choices. These studies also cannot make precise
mathematical predictions about the effect of deception or deception-detecting
techniques \cite{key-19}. Understanding deception in a quantitative
framework could help to give results rigor and predictability.

To achieve this rigor and predictability, we analyze deception through
the framework of game theory. This framework allows making quantitative,
verifiable predictions, and enables the study of situations involving
free choice (the option to deceive or not to deceive) and well-defined
incentives \cite{key-19}. Specifically, the area of incomplete information
games allows modeling the information asymmetry that forms part and
parcel of deception. In a signaling game, a receiver observes a piece
of private information and communicates a message to a receiver, who
chooses an action. The receiver's best action depends on his belief
about the private information of the sender. But the sender may use
strategies in which he conveys or does not convey this private information.
It is natural to make connections between the signaling game terminology
of pooling, separating, and partially-separating equilibria and deceptive,
truthful, and partially-truthful behavior. Thus, game theory provides
a suitable framework for studying deception. 

\begin{figure}

\begin{centering}
\includegraphics[width=0.8\columnwidth]{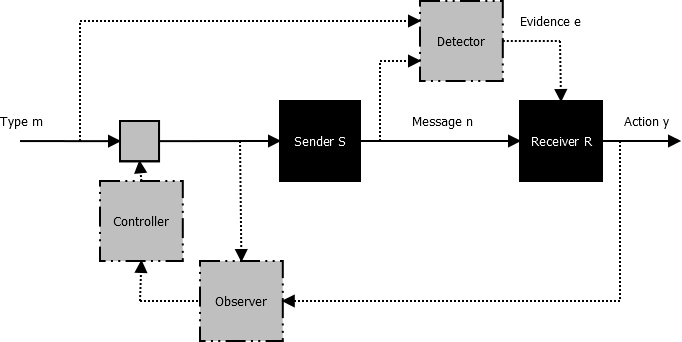}
\par\end{centering}

\protect\caption{\label{fig:A-general-framework}A general framework for mechanism
design. Manipulating the environment in which deception takes place
in a signaling game could include adding additional blocks as well
as manipulating exogenous parameters of the game. In general, type
$m$ can be manipulated by input from a \emph{controller} before reaching
the sender. The controller can rely on an \emph{observer} to estimate
unknown states. In this paper, we specifically study the roll of a
\emph{detector}, which compares type to message and emits evidence
for deception.}

\end{figure}

Beyond analyzing equilibria, we also want to design solutions that
control the environment in which deception takes place. This calls
for the reverse game theory perspective of \emph{mechanism design}.
In mechanism design, exogenous factors are manipulated in order to
design the outcome of a game. In signaling games, these solutions
might seek to obtain target utilities or a desired level of information
communication. If the deceiver in the signaling game has the role
of an adversary - for problems in security or privacy, for example
- a defender often wants to design methods to limit the amount of
deception. But defenders may also use deception to their advantage.
In this case, it is the adversary who may try to implement mechanisms
to mitigate the effects of the deception. A more general mechanism
design perspective for signaling games could consider other ways of
manipulating the environment, such as feedback and observation (Fig.
\ref{fig:A-general-framework}).

In this paper, we study deception in two different frameworks. The
first framework is a typical game of costless communication between
a sender and receiver known as \emph{cheap-talk}. In the second framework,
we add the element of deception detection, forming a game of \emph{cheap-talk
with evidence}. This latter model includes a move by nature after
the action of the sender, which yields evidence for deception with
some probability. In order provide a concrete example, we consider
a specific use of deception for defense, and the employment of antideceptive
techniques by an attacker. In this scenario, a defender uses honeypots
disguised as normal systems to protect a network, and an adversary
implements honeypot detection in order to strike back against this
deception. We give an example of how an adversary might obtain evidence
for deception through a timing classification known as \emph{fuzzy
benchmarking}. Finally, we show how network defenders need to bolster
their capabilities in order to maintain the same results in the face
of honeypot detection. This mechanism design approach reverses the
mappings from adversary power to evidence detection and evidence detection
to game outcome. Although we apply it to a specific research problem,
our approach is quite general and can be used in deceptive interactions
in both interpersonal deception and deception in cyber security. Our
main contributions include 1) developing a model for signaling games
with deception detection, and analyzing how this model includes traditional
signaling games and complete information games as special cases, 2)
demonstrating that the ability to detect deception causes pure strategy
equilibria to disappear under certain conditions, and 3) showing that
deception detection by an adversary could actually increase the utility
obtained by a network defender. These results have specific implications
for network defense through honeypot deployment, but can be applied
to a large class of strategic interactions involving deception in
both physical and cyberspace.

The rest of the paper proceeds as follows. Section \ref{sec:Cheap-Talk-Signaling-Games}
reviews cheap-talk signaling games and the solution concept of perfect
Bayesian Nash equilibrium. We use this framework to analyze the honeypot
scenario in Section \ref{sec:AnalysisNoEv}. Section \ref{sec:Cheap-Talk-Ev}
adds the element of deception detection to the signaling game. We
describe an example of how this detection might be implemented in
Section \ref{sec:Deception-Detection-Example}. Then we analyze the
resulting game in section \ref{sec:AnalysisEv}. In Section \ref{sec:Mechanism-Design},
we discuss a case study in which a network defender needs to change
in order to respond to the advent of honeypot detection. We review
related work in Section \ref{sec:Related-Work}, and conclude the
paper in Section \ref{sec:Discussion}.

\section{Cheap-Talk Signaling Games\label{sec:Cheap-Talk-Signaling-Games}}

In this section, we review the concept of signaling games, a class
of two-player, dynamic, incomplete information games. The information
asymmetry and dynamic nature of these games captures the essence of
deception, and the notion of separating, pooling, or partially-separating
equilibria can be related to truthful, deceptive, or partially-truthful
behavior.

\subsection{Game Model}

\begin{figure}
\begin{centering}
\includegraphics[width=0.67\columnwidth]{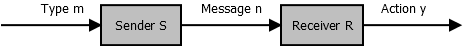}
\par\end{centering}

\protect\caption{\label{fig:Block-diagram-signal}Block diagram of a signaling game
with two discrete types, messages, and actions.}
\end{figure}

Our model consists of a signaling game in which the types, messages,
and actions are taken from discrete sets with two elements. Call this
two-player, incomplete information game $\mathcal{G}$. In $\mathcal{G}$,
a sender, $S$, observes a type $m\in M=\left\{ 0,1\right\} $ drawn
with probabilities $p\left(0\right)$ and $p\left(1\right)=1-p\left(0\right)$.
He then sends a message, $n\in N=\left\{ 0,1\right\} $ to the receiver,
$R$. After observing the message (but not the type), $R$ plays an
action $y\in Y=\left\{ 0,1\right\} $ . The flow of information between
sender and receiver is depicted in Fig. \ref{fig:Block-diagram-signal}.
Let $u^{S}\left(y,m\right)$ and $u^{R}\left(y,m\right)$ be the utility
obtained by $S$ and $R$, respectively, when the type is $m$ and
the receiver plays action $y$. Notice that the utilities are not
directly dependent on the message, $n$; hence the description of
this model as a ``cheap-talk'' game.

The sender's strategy consists of playing a message $n$, after observing
a type $m$, with probability $\sigma_{S}\left(n\,|\, m\right)$.
The receiver's strategy consists of playing an action $y$, after
observing a message $n$, with probability $\sigma_{R}\left(y\,|\, n\right)$.
Denote the sets of all such strategies as $\Gamma^{S}$, and $\Gamma^{R}$.
Define expected utilities for the sender and receiver as $U^{S}:\,\Gamma^{S}\times\Gamma^{R}\rightarrow\mathbb{R}$
and $U^{R}:\,\Gamma^{S}\times\Gamma^{R}\rightarrow\mathbb{R}$, such
that $U^{S}\left(\sigma_{S},\sigma_{R}\right)$ and $U^{R}\left(\sigma_{S},\sigma_{R}\right)$
are the expected utilities for the sender and receiver, respectively,
when the sender and receiver play according to the strategy profile
$\left(\sigma_{S},\sigma_{R}\right)$. Finally, define $\tilde{U}^{S}:\,\Gamma^{S}\times\Gamma^{R}\times M\to\mathbb{R}$
and $\tilde{U}^{R}:\,\Gamma^{R}\times M\times N\to\mathbb{R}$ such
that $\tilde{U}^{S}\left(\sigma_{S},\sigma_{R},m\right)$ gives the
expected utility for $S$ for playing $\sigma_{S}$ when $R$ plays
$\sigma_{R}$ and the type is $m$, and $\tilde{U}^{R}\left(\sigma_{R},m,n\right)$
gives the expected utility for $R$ for playing $\sigma_{R}$ when
the type is $m$ and she observes message $n$.

\subsection{Perfect Bayesian Nash Equilibrium}

We now review the concept of Perfect Bayesian Nash equilibrium, the
natural extension of subgame perfection to games of incomplete information.

A Perfect Bayesian Nash equilibrium (see \cite{key-16}) of signaling
game $\mathcal{G}$ is a strategy profile $\left(\sigma_{S},\sigma_{R}\right)$
and posterior beliefs $\mu_{R}(m\,|\, n)$ of the receiver about the
sender such that 

\begin{equation}
\forall m\in M,\,\sigma_{S}\in\underset{\bar{\sigma}_{S}\in\Gamma^{S}}{\arg\max\,}\tilde{U}^{S}\left(\bar{\sigma}_{S},\sigma_{R},m\right),\label{eq:PBE1}
\end{equation}

\begin{equation}
\forall n\in N,\,\sigma_{R}\in\underset{\bar{\sigma}_{R}\in\Gamma^{R}}{\arg\max}\,\underset{\bar{m}\in M}{\sum}\mu_{R}\left(\bar{m}\,|\, n\right)\tilde{U}^{R}\left(\bar{\sigma}_{R},\bar{m},n\right),
\end{equation}

\begin{equation}
\mu_{R}\left(m\,|\, n\right)=\begin{cases}
\frac{\sigma_{S}\left(n\,|\, m\right)p\left(m\right)}{\underset{\bar{m}\in M}{\sum}\sigma_{S}\left(n\,|\,\bar{m}\right)p\left(\bar{m}\right)}, & \text{if}\underset{\bar{m}\in M}{\sum}\sigma_{S}\left(n\,|\,\bar{m}\right)p\left(\bar{m}\right)>0\\
\text{any distrubution on }M, & \text{if}\underset{\bar{m}\in M}{\sum}\sigma_{S}\left(n\,|\,\bar{m}\right)p\left(\bar{m}\right)=0
\end{cases}.\label{eq:def1-beliefUp}
\end{equation}

Eq. \ref{eq:PBE1} requires $S$ to maximize his expected utility
for the strategy played by $R$ for all types $m$. The second equation
requires that, for all messages $n$, $R$ maximizes his expected
utility against the strategy played by $S$ given his beliefs. Finally,
Eq. \ref{eq:def1-beliefUp} requires the beliefs of $R$ about the
type to be consistent with the strategy played by $S$, using Bayes'
Law to update his prior belief according to $S$'s strategy.

\section{Analysis of Deceptive Conflict Using Signaling Games\label{sec:AnalysisNoEv}}

In this section, we describe an example of deception in cyber security
using signaling games. These type of models have been used, for instance,
in \cite{key-17,key-27,key-5,key-25}. We give results here primarily
in order to show how the results change after we add the factor of
evidence emission in Section \ref{sec:AnalysisEv}. 

\begin{figure}
\begin{centering}
\includegraphics[width=0.6\columnwidth]{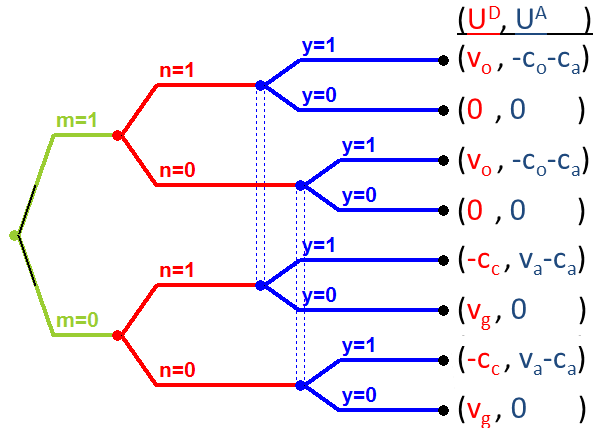}
\par\end{centering}

\protect\caption{\label{fig:ExtFormGHoney}Extensive form of $\mathcal{G}_{honey}$,
a game in which defender $S$ chooses whether to disguise systems
in a network of computers, and an attacker $R$ attempts to gain from
compromising normal systems but withdrawing from honeypots. Note that
the type $m$ is determined by a chance move.}
\end{figure}

Consider a game $\mathcal{G}_{honey}$, in which a defender uses honeypots
to protect a network of computers. We consider a model and parameters
from \cite{key-17}, with some adaptations. In this game, the ratio
of normal systems to honeypots is considered fixed. Based on this
ratio, nature assigns a \emph{type} - normal system or honeypot -
to each system in the network. The sender is the network defender,
who can choose to reveal the type of each system or disguise the systems.
He can disguise honeypots as normal systems and disguise normal systems
as honeypots. The \emph{message} is thus the network defender's portrayal
of the system. The receiver in this game is the attacker, who observes
the defender's portrayal of the system but not the actual type of
the system. He forms a \emph{belief} about the actual type of the
system given the sender's message, and then chooses an \emph{action:}
attack or withdraw%
\footnote{In the model description in \cite{key-17}, the attacker also has
an option to condition his attack on testing the system. We omit this
option, because we will consider the option to test the system through
a different approach in the signaling game with evidence emission
in Section \ref{sec:AnalysisEv}.%
}. Table \ref{tab:Parameters-GHoney} gives the parameters of $\mathcal{G}_{honey}$,
and the extensive form of $\mathcal{G}_{honey}$ is given in Fig.
\ref{fig:ExtFormGHoney}. We have used the game theory software \emph{Gambit}
\cite{key-6} for this illustration, as well as for simulating the
results of games later in the paper. 

\begin{table*}
\protect\caption{\label{tab:Parameters-GHoney}Parameters of $\mathcal{G}_{honey}$.
M.S. signifies Mixed Strategy}

\centering{}%
\begin{tabular}{|c|c|}
\hline 
Parameter Symbol & Meaning\tabularnewline
\hline 
\hline 
$S$ & Network defender\tabularnewline
\hline 
$R$ & Network attacker\tabularnewline
\hline 
$m\in\left\{ 0,1\right\} $ & Type of system ($0$: normal; $1$: honeypot)\tabularnewline
\hline 
$n\in\left\{ 0,1\right\} $ & Defender description of system ($0$: normal; $1$: honeypot)\tabularnewline
\hline 
$y\in\left\{ 0,1\right\} $ & Attacker action ($0$: withdraw; $1$: attack)\tabularnewline
\hline 
$p(m)$ & Prior probability of type $m$\tabularnewline
\hline 
$\sigma_{S}\left(n\,|\, m\right)$ & Sender MS prob. of describing type $m$ as $n$ \tabularnewline
\hline 
$\sigma_{R}\left(y\,|\, n\right)$ & Receiver MS prob. of action $y$ given description $n$ \tabularnewline
\hline 
$v_{o}$ & Defender benefit of observing attack on honeypot\tabularnewline
\hline 
$v_{g}$ & Defender benefit of avoiding attack on normal system\tabularnewline
\hline 
$-c_{c}$ & Defender cost of normal system being compromised \tabularnewline
\hline 
$v_{a}$ & Attacker benefit of comprimizing normal system\tabularnewline
\hline 
$-c_{a}$ & Attacker cost of attack on any type of system\tabularnewline
\hline 
$-c_{o}$ & Attacker additional cost of attacking honeypot\tabularnewline
\hline 
\end{tabular}
\end{table*}

In order to characterize the equilibria of $\mathcal{G}_{honey}$,
define two constants: $\mathcal{CB}_{0}^{R}$ and $\mathcal{CB}_{1}^{S}$.
Let $\mathcal{CB}_{0}^{R}$ give the relative benefit to $R$ for
playing attack ($y=1$) compared to playing withdraw ($y=0$) when
the system is a normal system ($m=0$), and let $\mathcal{CB}_{1}^{R}$
give the relative benefit to $R$ for playing withdraw compared to
playing attack when the system is a honeypot ($m=1$). These constants
are defined by Eq. \ref{eq:CB0} and Eq. \ref{eq:CB1}.

\begin{equation}
\mathcal{CB}_{0}^{R}\triangleq u^{R}\left(1,0\right)-u^{R}\left(0,0\right)\label{eq:CB0}
\end{equation}

\begin{equation}
\mathcal{CB}_{1}^{R}\triangleq u^{R}\left(0,1\right)-u^{R}\left(1,1\right)\label{eq:CB1}
\end{equation}

We now find the pure-strategy separating and pooling equilibria of
$\mathcal{G}_{honey}$. 
\begin{thm}
The equilibria of $\mathcal{G}_{honey}$ differ in form in three parameter
regions: \end{thm}
\begin{itemize}
\item Attack-favorable: $p\left(0\right)\mathcal{CB}_{0}^{R}>\left(1-p\left(0\right)\right)\mathcal{CB}_{1}^{R}$ 
\item Defend-favorable: \textbf{$p\left(0\right)\mathcal{CB}_{0}^{R}<\left(1-p\left(0\right)\right)\mathcal{CB}_{1}^{R}$}
\item Neither-favorable: \textbf{$p\left(0\right)\mathcal{CB}_{0}^{R}=\left(1-p\left(0\right)\right)\mathcal{CB}_{1}^{R}$}
\end{itemize}
In attack-favorable, $p\left(0\right)\mathcal{CB}_{0}^{R}>\left(1-p\left(0\right)\right)\mathcal{CB}_{1}^{R}$,
meaning loosely that the relative benefit to the receiver for attacking
normal systems is greater than the relative loss to the receiver for
attacking honeypots. In defend-favorable, \textbf{$p\left(0\right)\mathcal{CB}_{0}^{R}<\left(1-p\left(0\right)\right)\mathcal{CB}_{1}^{R}$},
meaning that the relative loss for attacking honeypots is greater
than the relative benefit from attacking normal systems. In neither-favorable,
\textbf{$p\left(0\right)\mathcal{CB}_{0}^{R}=\left(1-p\left(0\right)\right)\mathcal{CB}_{1}^{R}$}.
We omit analysis of the neither-favorable region because it only arises
with exact equality in the game parameters.

\subsection{Separating Equilibria}

In separating equilibria, the sender plays different pure strategies
for each type that he observes. Thus, he completely reveals the truth.
The attacker $R$ in $\mathcal{G}_{honey}$ wants to attack normal
systems but withdraw from honeypots. The defender $S$ wants the opposite:
that the attacker attack honeypots and withdraw from normal systems.
Thus, Theorem \ref{thm:No-separating-equilibria} should come as no
surprise.
\begin{thm}
\label{thm:No-separating-equilibria}No separating equilibria exist
in $\mathcal{G}_{honey}$.
\end{thm}

\subsection{Pooling Equilibria}

In pooling equilibria, the sender plays the same strategies for each
type. This is deceptive behavior because the sender's messages do
not convey the type that he observes. The receiver relies only on
prior beliefs about the distribution of types in order to choose his
action. Theorem \ref{thm:pooling-noEv-attack} gives the pooling equilibria
of $\mathcal{G}_{honey}$ in the attack-favorable region. 
\begin{thm}
\label{thm:pooling-noEv-attack}$\mathcal{G}_{honey}$ supports the
following pure strategy pooling equilibria in the attack-favorable
parameter region:
\begin{equation}
\forall m\in M,\,\sigma_{S}\left(1\,|\, m\right)=1,
\end{equation}
\begin{equation}
\forall n\in N,\,\sigma_{R}\left(1\,|\, n\right)=1,
\end{equation}
\begin{equation}
\mu_{R}\left(1\,|\,0\right)\leq\frac{\mathcal{CB}_{0}^{R}}{\mathcal{CB}_{0}^{R}+\mathcal{CB}_{1}^{R}};\;\mu_{R}\left(1\,|\,1\right)=p\left(1\right),
\end{equation}
and
\begin{equation}
\forall m\in M,\,\sigma_{S}\left(1\,|\, m\right)=0,
\end{equation}
\begin{equation}
\forall n\in N,\,\sigma_{R}\left(1\,|\, n\right)=1,
\end{equation}
\begin{equation}
\mu_{R}\left(1\,|\,0\right)=p\left(1\right);\;\mu_{R}\left(1\,|\,1\right)\leq\frac{\mathcal{CB}_{0}^{R}}{\mathcal{CB}_{0}^{R}+\mathcal{CB}_{1}^{R}},
\end{equation}
both with expected utilities given by

\begin{equation}
U^{S}\left(\sigma_{S},\sigma_{R}\right)=u^{S}\left(1,1\right)-p\left(0\right)\left(u^{S}\left(1,1\right)-u^{S}\left(1,0\right)\right),
\end{equation}

\begin{equation}
U^{R}\left(\sigma_{S},\sigma_{R}\right)=u^{R}\left(1,1\right)-p\left(0\right)\left(u^{R}\left(1,1\right)-u^{R}\left(1,0\right)\right).
\end{equation}

\end{thm}
Similarly, Theorem \ref{thm:pooling-noEv-defend} gives the pooling
equilibria of $\mathcal{G}_{honey}$ in the defend-favorable region.
\begin{thm}
\label{thm:pooling-noEv-defend}$\mathcal{G}_{honey}$ supports the
following pure strategy pooling equilibria in the defend-favorable
parameter region:
\begin{equation}
\forall m\in M,\,\sigma_{S}\left(1\,|\, m\right)=1,
\end{equation}
\begin{equation}
\forall n\in N,\,\sigma_{R}\left(1\,|\, n\right)=0,
\end{equation}
\begin{equation}
\mu_{R}\left(1\,|\,0\right)\geq\frac{\mathcal{CB}_{0}^{R}}{\mathcal{CB}_{0}^{R}+\mathcal{CB}_{1}^{R}};\;\mu_{R}\left(1\,|\,1\right)=p\left(1\right),
\end{equation}
and
\begin{equation}
\forall m\in M,\,\sigma_{S}\left(1\,|\, m\right)=0,
\end{equation}
\begin{equation}
\forall n\in N,\,\sigma_{R}\left(1\,|\, n\right)=0,
\end{equation}
\begin{equation}
\mu_{R}\left(1\,|\,0\right)=p\left(1\right);\;\mu_{R}\left(1\,|\,1\right)\geq\frac{\mathcal{CB}_{0}^{R}}{\mathcal{CB}_{0}^{R}+\mathcal{CB}_{1}^{R}},
\end{equation}
both with expected utilities given by

\begin{equation}
U^{S}\left(\sigma_{S},\sigma_{R}\right)=p\left(0\right)\left(u^{S}\left(0,0\right)-u^{S}\left(0,1\right)\right)+u^{S}\left(0,1\right),
\end{equation}

\begin{equation}
U^{R}\left(\sigma_{S},\sigma_{R}\right)=p\left(0\right)\left(u^{R}\left(0,0\right)-u^{R}\left(0,1\right)\right)+u^{R}\left(0,1\right).
\end{equation}

\end{thm}
In both cases, it is irrelevant whether the defender always sends
$1$ or always sends $0$ (always describes systems as honeypots or
always describes systems as normal systems); the effect is that the
attacker ignores the description. In the attack-favorable region,
the attacker always attacks. In the defend-favorable region, the attacker
always withdraws.

\subsection{Discussion of $\mathcal{G}_{honey}$ Equilibria}

\begin{figure}

\begin{centering}
\includegraphics[width=0.6\columnwidth]{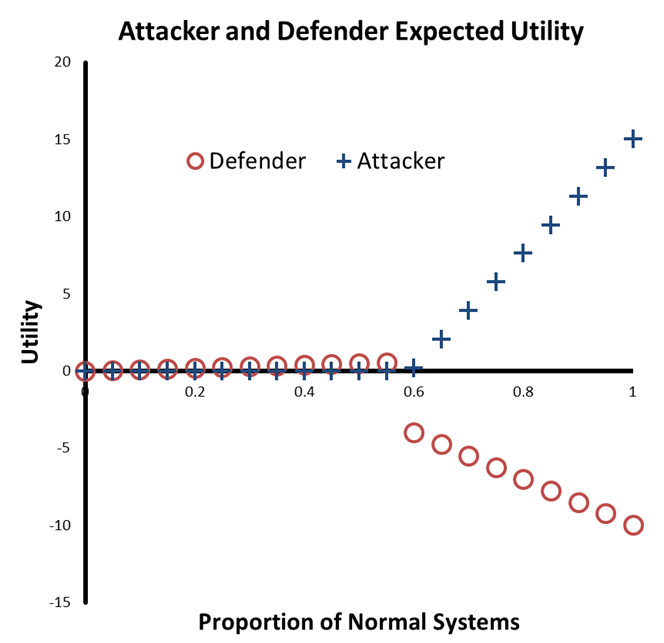}
\par\end{centering}

\protect\caption{\label{fig:Expected-Utilities-NoEv}Expected Utilities verses Fraction
of Normal Systems in Network.}

\end{figure}

We will discuss these equilibria more when we compare them with the
equilibria of the game with evidence emission. Still, we note one
aspect of the equilibria here. At \textbf{$p\left(0\right)\mathcal{CB}_{0}^{R}=\left(1-p\left(0\right)\right)\mathcal{CB}_{1}^{R}$},
the expected utility is continuous for the receiver, but not for the
sender. As shown in Fig. \ref{fig:Expected-Utilities-NoEv}, the sender's
(network defender's) utility sharply improves if he transitions from
having \textbf{$p\left(0\right)\mathcal{CB}_{0}^{R}>\left(1-p\left(0\right)\right)\mathcal{CB}_{1}^{R}$
}to\textbf{ $p\left(0\right)\mathcal{CB}_{0}^{R}<\left(1-p\left(0\right)\right)\mathcal{CB}_{1}^{R}$}\emph{,
i.e.} from having\emph{ }$40\%$ honeypots to having $41\%$ honeypots.
This is an obvious mechanism design consideration. We will analyze
this case further in the section on mechanism design.

\section{Cheap-Talk Signaling Games with Evidence\label{sec:Cheap-Talk-Ev}}

In Section \ref{sec:AnalysisNoEv}, we used a typical signaling game
to model deception in cyberspace (in $\mathcal{G}_{honey}$). In this
section, we add to this game the possibility that the sender gives
away evidence of deception.

In a standard signaling game, the receiver's belief about the type
is based only on the messages that the sender communicates and his
prior belief. In many deceptive interactions, however, there is some
probability that the sender gives off evidence of deceptive behavior.
In this case, the receiver's beliefs about the sender's private information
may be updated both based upon the message of the sender and by evidence
of deception.

\subsection{Game Model \label{sub:ProtocolEvEmiss}}

\begin{figure}
\begin{centering}
\includegraphics[width=0.67\columnwidth]{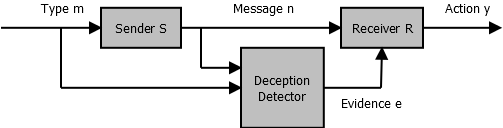}
\par\end{centering}

\protect\caption{\label{fig:BlockWithEvidence}Block diagram of a signaling game with
evidence emission.}
\end{figure}

Let $\mathcal{G}^{evidence}$ denote a signaling game with belief
updating based both on sender message and on evidence of deception.
This game consists of four steps, in which step 3 is new:
\begin{enumerate}
\item Sender, $S$, observes type, $m\in M=\left\{ 0,1\right\} $.
\item $S$ communicates a message, $n\in N=\left\{ 0,1\right\} $, chosen
according to a strategy $\sigma_{S}\left(n\,|\, m\right)\in\Gamma^{S}=\Delta N$
based on the type $m$ that he observes.
\item \emph{$S$ emits evidence, $e\in E=\left\{ 0,1\right\} $ with probability
$\lambda\left(e\,|\, m,n\right)$. Signal $e=1$ represents evidence
of deception and $e=0$ represents no evidence of deception. }
\item Receiver $R$ responds with an action, $y\in Y=\left\{ 0,1\right\} $,
chosen according to a strategy $\sigma_{R}\left(y\,|\, n,e\right)\in\Gamma^{R}=\Delta Y$
based on the message $n$ that he receives and evidence $e$ that
he observes.
\item $S$, $R$ receive $u^{S}\left(y,m\right)$, $u^{R}\left(y,m\right)$. 
\end{enumerate}
Evidence $e$ is another signal that is available to $R$, in addition
to the message $n$. This signal could come, \emph{e.g.}, from a \emph{detector},
which generates evidence with a probability that is a function of
$m$ and $n$. The detector implements the function $\lambda\left(e\,|\, m,n\right)$.
We depict this view of the signaling game with evidence emission in
Fig. \ref{fig:BlockWithEvidence}. We assume that $\lambda\left(e\,|\, m,n\right)$
is common knowledge to both the sender and receiver. Since evidence
is emitted with some probability, we model this as a move by a ``chance''
player, just as we model the random selection of the type at the beginning
of the game as a move by a chance player. The outcome of the new chance
move will be used by $R$ together with his observation of $S$'s
action to formulate his belief about the type $m$. We describe this
belief updating in the next section.

\subsection{Two-step Bayesian Updating}

Bayesian updating is a two-step process, in which the receiver first
updates his belief about the type based on the observed message of
the sender, and then updates his belief a second time based on the
evidence emitted. The following steps formulate the update process.
\begin{enumerate}
\item $R$ observes $S$'s action. He computes belief $\mu_{R}\left(m\,|\, n\right)$
based on the prior likelihoods $p\left(m\right)$ of each type and
$S$'s message $n$ according to Eq. \ref{eq:def1-beliefUp}, which
we rewrite here in Eq. \ref{eq:firstStepUpdateBelief-1}. 
\begin{equation}
\mu_{R}\left(m\,|\, n\right)=\begin{cases}
\frac{\sigma_{S}\left(n\,|\, m\right)p\left(m\right)}{\underset{\bar{m}\in M}{\sum}\sigma_{S}\left(n\,|\,\bar{m}\right)p\left(\bar{m}\right)}, & \text{if}\underset{\bar{m}\in M}{\sum}\sigma_{S}\left(n\,|\,\bar{m}\right)p\left(\bar{m}\right)>0\\
\text{any distribution on }M, & \text{if}\underset{\bar{m}\in M}{\sum}\sigma_{S}\left(n\,|\,\bar{m}\right)p\left(\bar{m}\right)=0
\end{cases}\label{eq:firstStepUpdateBelief-1}
\end{equation}

\item $S$ computes a new belief based on the evidence emitted. The prior
belief in this second step is given by $\mu_{R}\left(m\,|\, n\right)$
obtained in the first step. The conditional probability of emitting
evidence $e$ when the type is $m$ and the sender communicates message
$n$ is $\lambda\left(e\,|\, m,n\right)$. Thus, the receiver updates
his belief in this second step according to
\begin{equation}
\mu_{R}\left(m\,|\, n,e\right)=\begin{cases}
\frac{\lambda\left(e\,|\, m,n\right)\mu_{R}\left(m\,|\, n\right)}{\underset{\bar{m}\in M}{\sum}\lambda\left(e\,|\,\bar{m},n\right)\mu_{R}\left(\bar{m}\,|\, n\right)}, & \text{if}\underset{\bar{m}\in M}{\sum}\lambda\left(e\,|\,\bar{m},n\right)\mu_{R}\left(\bar{m}\,|\, n\right)>0\\
\text{any distribution on }M, & \text{if}\underset{\bar{m}\in M}{\sum}\lambda\left(e\,|\,\bar{m},n\right)\mu_{R}\left(\bar{m}\,|\, n\right)=0
\end{cases}.\label{eq:secondStepUpdateBelief-1}
\end{equation}

\end{enumerate}
Having formulated the belief updating rules, we now give the conditions
for a Perfect Bayesian Nash equilibrium in our signaling game with
evidence emission.

\subsection{Perfect Bayesian Nash Equilibrium in Signaling Game with Evidence}

The conditions for a Perfect Bayesian Nash Equilibrium of our augmented
game are the same as those for the original signaling game, except
that the belief update includes the use of emitted evidence. Here,
however, we must also define a new utility function for $R$ that
takes expectation conditional upon $e$ in addition to $n$. Define
this utility function by $\hat{U}^{R}:\,\Gamma^{R}\times M\times N\times E\to\mathbb{R}$
such that $\hat{U}^{R}\left(\sigma_{R},m,n,e\right)$ gives the expected
utility for $R$ for playing $\sigma_{R}$ when the type is $m$ and
she observes message $n$ and evidence $e$.
\begin{defn}
A perfect Bayesian Nash equilibrium of the game $\mathcal{G}^{evidence}$
is a strategy profile $\left(\sigma_{S},\sigma_{R}\right)$ and posterior
beliefs $\mu_{R}(m\,|\, n,e)$, such that system given by Eq. \ref{eq:senderMaxEv}
through Eq. \ref{eq:beliefStep2Thm} are simultaneously satisfied.

\begin{equation}
\forall m\in M,\,\sigma_{S}\in\underset{\bar{\sigma}_{S}\in\Gamma^{S}}{\arg\max\,}\tilde{U}^{S}\left(\bar{\sigma}_{S},\sigma_{R},m\right)\label{eq:senderMaxEv}
\end{equation}

\begin{equation}
\forall n\in N,\,\forall e\in E,\,\sigma_{R}\in\underset{\bar{\sigma}_{R}\in\Gamma^{R}}{\arg\max\,}\underset{\bar{m}\in M}{\sum}\mu_{R}(\bar{m}\,|\, n,e)\hat{U}^{R}\left(\bar{\sigma}_{R},\bar{m},n,e\right)\label{eq:receiverMaxEv}
\end{equation}
\begin{equation}
\forall n\in N,\,\mu_{R}\left(m\,|\, n\right)=\begin{cases}
\frac{\sigma_{S}\left(n\,|\, m\right)p\left(m\right)}{\underset{\bar{m}\in M}{\sum}\sigma_{S}\left(n\,|\,\bar{m}\right)p\left(\bar{m}\right)}, & \text{if}\underset{\bar{m}\in M}{\sum}\sigma_{S}\left(n\,|\,\bar{m}\right)p\left(\bar{m}\right)>0\\
\text{any distribution on }M, & \text{if}\underset{\bar{m}\in M}{\sum}\sigma_{S}\left(n\,|\,\bar{m}\right)p\left(\bar{m}\right)=0
\end{cases}\label{eq:beliefStep1Thm}
\end{equation}

\begin{equation}
\begin{array}{c}
\forall n\in N,\,\forall e\in E,\\
\mu_{R}\left(m\,|\, n,e\right)=
\end{array}\begin{cases}
\frac{\lambda\left(e\,|\, m,n\right)\mu_{R}\left(m\,|\, n\right)}{\underset{\bar{m}\in M}{\sum}\lambda\left(e\,|\,\bar{m},n\right)\mu_{R}\left(\bar{m}\,|\, n\right)}, & \text{if}\underset{\bar{m}\in M}{\sum}\lambda\left(e\,|\,\bar{m},n\right)\mu_{R}\left(\bar{m}\,|\, n\right)>0\\
\text{any distribution on }M, & \text{if}\underset{\bar{m}\in M}{\sum}\lambda\left(e\,|\,\bar{m},n\right)\mu_{R}\left(\bar{m}\,|\, n\right)=0
\end{cases}\label{eq:beliefStep2Thm}
\end{equation}

\end{defn}
Again, the first two definitions require the sender and receiver to
maximize their expected utilities. The third and fourth equations
require belief consistency in terms of Bayes' Law.

\section{Deception Detection Example in Network Defense\label{sec:Deception-Detection-Example}}

Consider again our example of deception in cyberspace in which a defender
protects a network of computer systems using honeypots. The defender
has the ability to disguise normal systems as honeypots and honeypots
as normal systems. In Section \ref{sec:AnalysisNoEv}, we modeled
this deception as if it were possible for the defender to disguise
the systems without any evidence of deception. In reality, attackers
may try to detect honeypots. For example, \emph{send-safe.com}'s ``Honeypot
Hunter'' \cite{key-22} checks lists of HTTPS and SOCKS proxies and
outputs text files of valid proxies, failed proxies, and honeypots.
It performs a set of tests which include opening a false mail server
on the local system to test the proxy connection, connecting to the
proxy port, and attempting to proxy back to its false mail server
\cite{key-21}. 

Another approach to detecting honeypots is based on timing. \cite{key-23}
used a process termed \emph{fuzzy benchmarking} in order to classify
systems as real machines or virtual machines, which could be used
\emph{e.g.}, as honeypots\emph{. }In this process, the authors run
a set of instructions which yield different timing results on different
host hardware architectures in order to learn more about the hardware
of the host system. Then, they run a loop of control modifying CPU
instructions (read and write control register 3, which induces a translation
lookaside buffer flush) that results in increased run-time on a virtual
machine compared to a real machine. The degree to which the run-times
are different between the real and virtual machines depends on the
number of sensitive instructions in the loop. The goal is to run enough
sensitive instructions to make the divergence in run-time - even in
the presence of internet noise - large enough to reliably classify
the system using a timing threshold. They do not identify limits to
the number of sensitive instructions to run, but we can imagine that
the honeypot detector might itself want to go undetected by the honeypot
and so might want to limit the number of instructions.

\begin{figure}
\begin{centering}
\includegraphics[width=0.67\columnwidth]{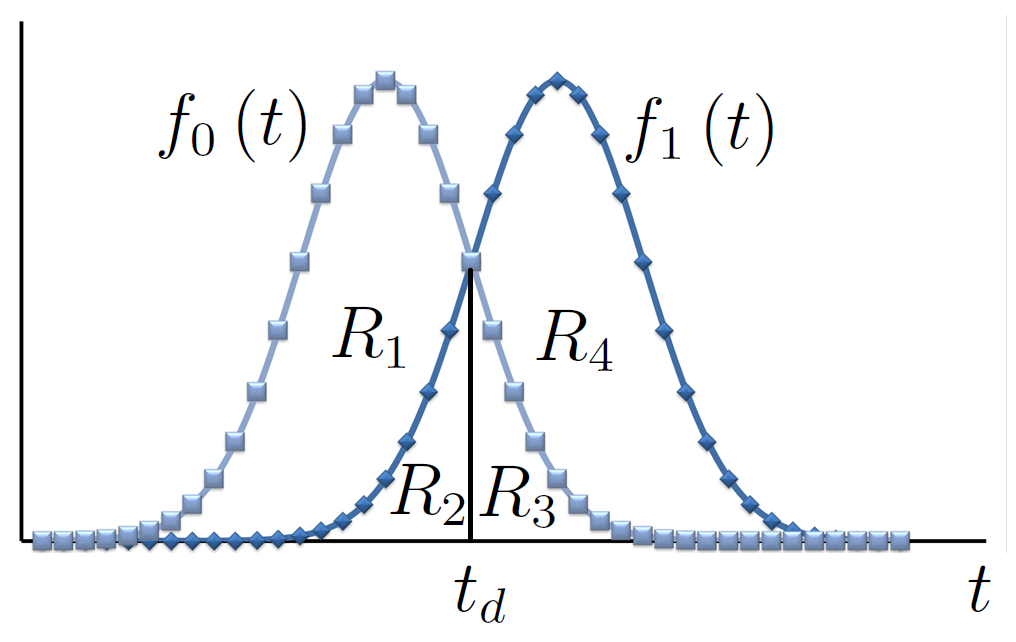}
\par\end{centering}

\protect\caption{\label{fig:ClassificationFuzzy}Classification of systems as normal
or virtual (\emph{e.g.} a honeypot) based on run-time for a set of
control modifying CPU instructions (based on fuzzy benchmarking in
\cite{key-23}).}

\end{figure}

Although they do not recount the statistical details, such an approach
could result in a classification problem which can only be accomplished
successfully with some probability. In Fig. \ref{fig:ClassificationFuzzy},
$t$ represents the execution time of the fuzzy benchmarking code.
The curve $f_{0}\left(t\right)$ represents the probability density
function for execution time for normal systems ($m=0$), and the curve
$f_{1}\left(t\right)$ represents the probability density function
for execution time for virtual machines ($m=1$). The execution time
$t_{d}$ represents a threshold time used to classify the system under
test. Let $AR_{i}$, $i\in\left\{ 1,2,3,4\right\} $ denote the area
under regions $R_{1}$ through $R_{4}$. We have defined $\lambda\left(e\,|\, m,n\right)$
to be the likelihood with which a system of type $m$ represented
as a system as type $n$ gives off evidence for deception $e$ (where
$e=1$ represents evidence for deception and $e=0$ represents evidence
for truth-telling). A virtual machine disguised as a normal system
may give off evidence for deception, in this case in terms of the
run-time of fuzzy benchmarking code. We would then have that

\begin{equation}
\begin{array}{c}
\lambda\left(1\,|\,1,0\right)=AR_{3}+AR_{4}\\
\lambda\left(0\,|\,1,0\right)=AR_{2}=1-\left(AR_{3}+AR_{4}\right)
\end{array}.
\end{equation}

If the system under test were actually a normal system, then the same
test could result in some likelihood of a false-positive result for
deception. Then, we would have

\begin{equation}
\begin{array}{c}
\lambda\left(1\,|\,0,0\right)=AR_{3}\\
\lambda\left(0\,|\,0,0\right)=AR_{1}+AR_{2}=1-\left(AR_{3}\right)
\end{array}.
\end{equation}

Let us assume that the likelihood with which one type of system masquerading
as another can be successfully detected is the same regardless of
whether it is a honeypot that is disguised as a normal system or it
is a normal system that is disguised as a honeypot. Denote this probability
as $\epsilon\in\left[0,1\right]$. Let $\delta\in\left[0,1\right]$
be defined as the likelihood of falsely detecting deception%
\footnote{Note that we assume that $\epsilon$ and $\delta$ are common knowledge;
the defender also knows the power of the adversary.%
}. These probabilities are given by

\begin{equation}
\epsilon=\lambda\left(1\,|\, m,n\right),\, m\neq n,\label{eq:epsilon}
\end{equation}

\begin{equation}
\delta=\lambda\left(1\,|\, m,n\right)\, m=n.\label{eq:delta}
\end{equation}
In \cite{key-23}, the authors tune the number of instructions for
the CPU to run in order to sufficiently differentiate normal systems
and honeypots. In this case, $\epsilon$ and \emph{$\delta$ }may
relate to the number of instructions that the detector asks the CPU
to run. In general, though, the factors which influence $\epsilon$
and \emph{$\delta$} could vary. Powerful attackers will have relatively
high $\epsilon$ and low $\delta$ compared to less powerful attackers.
Next, we study this network defense example using our model of signaling
games with evidence.

\section{Analysis of Network Defense using Signaling Games with Evidence\label{sec:AnalysisEv}}

\begin{figure}
\begin{centering}
\includegraphics[width=0.75\columnwidth]{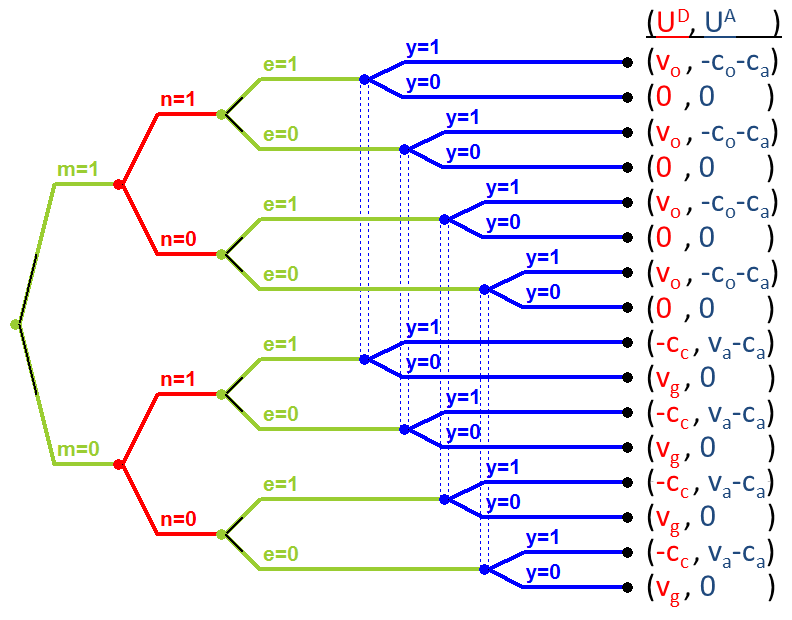}
\par\end{centering}

\protect\caption{\label{fig:Extensive-formGHEv}Extensive form depiction of $\mathcal{G}_{honey}^{evidence}$.
Note that the type $m$ and the evidence $e$ are both determined
by chance moves.}

\end{figure}

Figure \ref{fig:Extensive-formGHEv} depicts an extensive-form of
the signaling game with evidence for our network defense problem.
Call this game $\mathcal{G}_{honey}^{evidence}$. (See \cite{key-17}
for a more detailed explanation of the meaning of the parameters.)
In the extremes of $\epsilon$ and $\delta$, we will see that the
game degenerates into simpler types of games. 

\begin{figure}
\begin{centering}
\includegraphics[width=0.6\columnwidth]{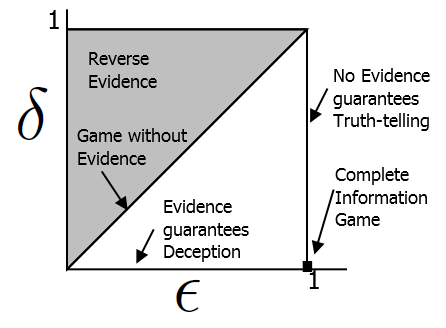}
\par\end{centering}

\protect\caption{\label{fig:Degenerate-cases}Degenerate cases of $\mathcal{G}_{honey}^{evidence}$}

\end{figure}

First, because $R$ updates his belief based on evidence emission
in a Bayesian manner, any situation in which $\delta=\epsilon$ will
render the evidence useless. The condition $\delta=\epsilon$ would
arise from an attacker completely powerless to detect deception. This
is indicated in Fig. \ref{fig:Degenerate-cases} by the region \emph{game
without evidence}, which we term $\mathcal{R}_{Weak}$ to indicate
an attacker with weak detection capability. 

Second, on the other extreme, we have the condition $\epsilon=1,\,\delta=0$,
which indicates that the attacker can always detect deception and
never registers false positives. Denote this region $\mathcal{R}_{Omnipotent}$
to indicate an attacker with omnipotent detection capability. $\mathcal{R}_{Omnipotent}$
degenerates into a \emph{complete information game} in which both
$S$ and $R$ are able to observe the type $m$. 

Third, we have a condition in which the attacker's detection capability
is such that \emph{evidence guarantees deception }(when $\delta=0$
but $\epsilon$ is not necessarily $1$) and a condition in which
the attacker's power is such that \emph{no evidence guarantees truth-telling}
(when $\epsilon=1$ but $\delta$ is not necessarily $0$). We can
term these two regions $\mathcal{R}_{Conservative}$ and $\mathcal{R}_{Aggressive}$,
because the attacker never detects a false positive in $\mathcal{R}_{Conservative}$
and never misses a sign for deception in $\mathcal{R}_{Aggressive}$. 

Finally, we have the region $\mathcal{R}_{Intermediate}$ in which
the attacker's detection capability is powerful enough that he correctly
detects deception with greater rate than he registers false positives,
but does not achieve $\delta=0$ or $\epsilon=1$. We list these attacker
conditions in Table \ref{degenerateCases}%
\footnote{We have defined these degenerate cases only for the case in which
$\epsilon\geq\delta$ - \emph{i.e.}, evidence for deception is more
likely to be emitted when the sender lies then when he tells the truth.
Mathematically, the equilibria of the game are actually symmetric
around the diagonal $\epsilon=\delta$ in Fig. \ref{fig:Degenerate-cases}.
This can be explained intuitively by considering the evidence emitted
to be ``evidence for truth-revelation'' in the upper-left corner.
In interpersonal deception, evidence for truth-revelation could correlate,
\emph{e.g.}, in the amount of spatial detail in a subject's account
of an event.%
}. Let us examine the equilibria of $\mathcal{G}_{honey}^{evidence}$
in these different cases. 

\begin{table}
\protect\caption{\label{degenerateCases}Attacker capabilities for degenerate cases
of $\mathcal{G}_{honey}^{evidence}$}

\centering{}%
\begin{tabular}{|c|c|c|}
\hline 
Name of Region & Description of Region & Parameter Values\tabularnewline
\hline 
\hline 
$\mathcal{R}_{Weak}$  & Game without evidence & $\delta=\epsilon$\tabularnewline
\hline 
$\mathcal{R}_{Omnipotent}$  & Complete information game & $\epsilon=1,\,\delta=0$\tabularnewline
\hline 
$\mathcal{R}_{Conservative}$  & Evidence guarantees deception & $\delta=0$ \tabularnewline
\hline 
$\mathcal{R}_{Aggressive}$  & No evidence guarantees truth-telling & $\epsilon=1$ \tabularnewline
\hline 
$\mathcal{R}_{Intermediate}$ & No guarantees & $\epsilon\neq1>\delta\neq0$\tabularnewline
\hline 
\end{tabular}
\end{table}

\subsection{Equilibria for $\mathcal{R}_{Weak}$ }

The equilibria for $\mathcal{R}_{Weak}$ are given by our analysis
of the game without evidence ($\mathcal{G}_{honey}$) in Section \ref{sec:AnalysisNoEv}.
Recall that a separating equilibrium was not sustainable, while pooling
equilibria did exist. Also, the equilibrium solutions fell into two
different parameter regions. The sender's utility was discontinuous
at the interface between parameter regions, creating an optimal proportion
of normal systems that could be included in a network while still
deterring attacks.

\subsection{Equilibria for $\mathcal{R}_{Omnipotent}$ }

For $\mathcal{R}_{Omnipotent}$, the attacker knows with certainty
the type of system (normal or honeypot) that he is facing. If the
evidence indicates that the system is a normal system, then he attacks.
If the evidence indicates that the system is a honeypot, then he withdraws.
The defender's description is unable to disguise the type of the system.
Theorem \ref{thm:pooling-Ev-Omni} gives the equilibrium strategies
and utilities.
\begin{thm}
\label{thm:pooling-Ev-Omni}$\mathcal{G}_{honey}^{evidence}$, under
adversary capabilities $\mathcal{R}_{Omnipotent}$ supports the following
equilibria:
\begin{equation}
\sigma_{S}\left(m\,|\, n\right)\in\Gamma^{S}
\end{equation}

\begin{equation}
\sigma_{R}\left(1\,|\, n,e\right)=\begin{cases}
n, & e=1\\
1-n, & e=0
\end{cases},\,\forall n\in N,
\end{equation}

\begin{equation}
\mu_{R}\left(1\,|\, n,e\right)=\begin{cases}
1-n, & e=1\\
n, & e=0
\end{cases},\,\forall n\in N,
\end{equation}

with expected utilities given by

\begin{equation}
U^{S}\left(\sigma_{S},\sigma_{R}\right)=p\left(0\right)\left(u^{S}\left(1,0\right)-u^{S}\left(0,1\right)\right)+u^{S}\left(0,1\right),
\end{equation}

\begin{equation}
U^{R}\left(\sigma_{S},\sigma_{R}\right)=p\left(0\right)\left(u^{R}\left(1,0\right)-u^{R}\left(0,1\right)\right)+u^{R}\left(0,1\right).
\end{equation}

\end{thm}
Similarly to $\mathcal{R}_{Weak}$, in $\mathcal{R}_{Omnipotent}$
the expected utilities for $S$ and $R$ are the same regardless of
the equilibrium strategy chosen (although the equilibrium strategy
profiles are not as interesting here because of the singular role
of evidence).

Next, we analyze the equilibria in the non-degenerate cases, $\mathcal{R}_{Conservative}$,
$\mathcal{R}_{Aggressive}$, and $\mathcal{R}_{Intermediate}$ , by
numerically solving for equilibria under selected parameter settings.

\subsection{Equilibria for $\mathcal{R}_{Conservative}$ , $\mathcal{R}_{Aggressive}$,
and $\mathcal{R}_{Intermediate}$ }

In Section \ref{sec:AnalysisNoEv}, we found analytical solutions
for the equilibria of a signaling game in which the receiver does
not have the capability to detect deception. In this section, we give
results concerning signaling games in which the receiver does have
the capability to detect deception, using illustrative examples rather
than an analytical solution. To study equilibria under the three non-degenerate
cases, we choose a set of parameters for the attacker and defender
utilities (Table \ref{tab:Sample-parameters-which}). In this model
(from \cite{key-17}), the defender gains utility from maintaining
normal systems that are not attacked in the network, and also from
observing attacks on honeypots. The defender incurs a loss if a normal
system is attacked. The attacker, on the other hand, gains only from
attacking a normal system; he incurs losses if he attacks a honeypot.

\begin{table}
\protect\caption{\label{tab:Sample-parameters-which}Sample parameters which describe
$\mathcal{GS}_{honey}^{evidence}$}

\centering{}%
\begin{tabular}{|c|c|}
\hline 
Parameter Symbol & Value\tabularnewline
\hline 
\hline 
$v_{o}$, sender utility from observing attack on honeypot & $5$\tabularnewline
\hline 
$v_{g}$, sender utility from normal system surviving & $1$\tabularnewline
\hline 
$-c_{C}$, sender cost for compromised normal system & $-10$\tabularnewline
\hline 
$-c_{o}-c_{a}$, cost due to attacker for attacking honeypot & $-22$\tabularnewline
\hline 
$0$, utility for attacker for withdrawing from any system & $0$\tabularnewline
\hline 
$v_{a}-c_{a}$, benefit of attacker for compromising normal system & $15$\tabularnewline
\hline 
\end{tabular}
\end{table}

Based on these parameters, we can find the equilibrium utilities at
each terminal node of Fig. \ref{fig:Extensive-formGHEv}. We study
examples in the attacker capability regions of $\mathcal{R}_{Conservative}$
, $\mathcal{R}_{Aggressive}$, and $\mathcal{R}_{Intermediate}$%
\footnote{The values of $\epsilon$ and $\delta$ are constrained by Table \ref{degenerateCases}.
Where the values are not set by the region, we choose them arbitrarily.
Specifically, we choose for $\mathcal{R}_{Weak}$, $\epsilon=0,\,\delta=0$;
for $\mathcal{R}_{Intermediate}$, $\epsilon=0.8,\,\delta=0.5$; for
$\mathcal{R}_{Conservative}$, $\epsilon=0.8,\,\delta=0$; for $\mathcal{R}_{Aggressive}$,
$\epsilon=1,\,\delta=0.5$, and for $\mathcal{R}_{Omnipotent}$, $\epsilon=1.0,\,\delta=0.$%
}. For each of these attacker capabilities, we look for equilibria
in pure strategies under three different selected values for the percentage
of normal systems (compared to honeypots) that make up a network.
For the high case, we set the ratio of normal systems to total systems
to be $p\left(0\right)=0.9$. Denote this case \emph{normal-saturated.
}For the medium case, we set $p\left(0\right)=0.6$. Denote this case
\emph{non-saturated}. Finally, label the low case, in which $p\left(0\right)=0.2$,
\emph{honeypot-saturated}. For comparison, we also include the equilibria
under the same game with no evidence emission (which corresponds to
$\mathcal{R}_{Weak}$ ), and the equilibria under the same game with
evidence that has a true-positive rate of $1.0$ and a false-positive
rate of $0$ (which corresponds to $\mathcal{R}_{Omnipotent}$ ).
In Table \ref{tab:Equilibria-for-Selected}, we list whether each
parameter set yields pure strategy equilibria. 
\begin{table*}
\protect\caption{\label{tab:Equilibria-for-Selected}Equilibria for Selected Parameter
Values in $\mathcal{R}_{Conservative}$ , $\mathcal{R}_{Aggressive}$,
and $\mathcal{R}_{Intermediate}$ , when the percentage of honeypots
in a network is high, medium, and low.}

\centering{}%
\begin{tabular}{|c|c|c|c|}
\hline 
{\small{}Saturation} & {\small{}$\mathcal{R}_{Weak}$ } & {\small{}$\mathcal{R}_{Intermediate}$, $\mathcal{R}_{Conservative}$,
$\mathcal{R}_{Aggressive}$} & {\small{}$\mathcal{R}_{Omnipotent}$ }\tabularnewline
\hline 
\emph{\small{}Normal} & {\small{}Yes} & {\small{}Yes} & {\small{}Yes}\tabularnewline
\hline 
\emph{\small{}None} & {\small{}Yes} & {\small{}None} & {\small{}Yes}\tabularnewline
\hline 
\emph{\small{}Honeypot} & {\small{}Yes} & {\small{}Yes} & {\small{}Yes}\tabularnewline
\hline 
\end{tabular}
\end{table*}

For adversary detection capabilities represented by $\mathcal{R}_{Weak}$
, we have a standard signaling game, and thus the well-known result
that a (pooling) equilibrium always exists. In $\mathcal{R}_{Omnipotent}$,
the deception detection is fool-proof, and thus the receiver knows
the type with certainty. We are left with a complete information game.
Essentially, the type merely determines which Stackelberg game the
sender and receiver play. Because pure strategy equilibria always
exist in Stackelberg games, $\mathcal{R}_{Omnipotent}$ also always
has pure-strategy equilibria. The rather unintuitive result comes
from $\mathcal{R}_{Intermediate}$, $\mathcal{R}_{Conservative}$,
and $\mathcal{R}_{Aggressive}$. In these ranges, the receiver's ability
to detect deception falls somewhere between no capability ($\mathcal{R}_{Weak}$
) and perfect capability ($\mathcal{R}_{Omnipotent}$ ). Those regions
exhibit pure-strategy equilibria, but the intermediate regions may
not. Specifically, they appear to fail to support pure-strategy equilibria
when the ratio of honeypots within the network does not fall close
to either $1$ or $0$. In Section \ref{sec:Mechanism-Design} on
mechanism design, we will see that this region plays an important
role in the comparison of network defense - and deceptive interactions
in general - with and without the technology for detecting deception.

\section{Mechanism Design for Detecting or Leveraging Deception\label{sec:Mechanism-Design}}

In this section, we discuss design considerations for a defender who
is protecting a network of computers using honeypots. In order to
do this, we choose a particular case study, and analyze how the network
defender can best set parameters to achieve his goals. We also discuss
the scenario from the point of view of the attacker. Specifically,
we examine how the defender can set the exogenous properties of the
interaction in 1) the case in which honeypots cannot be detected,
and 2) the case in which the attacker has implemented a method for
detecting honeypots. Then, we discuss the difference between these
two situations.

\subsection{Attacker Incapable of Honeypot Detection}

\begin{figure}

\begin{centering}
\includegraphics[width=0.6\columnwidth]{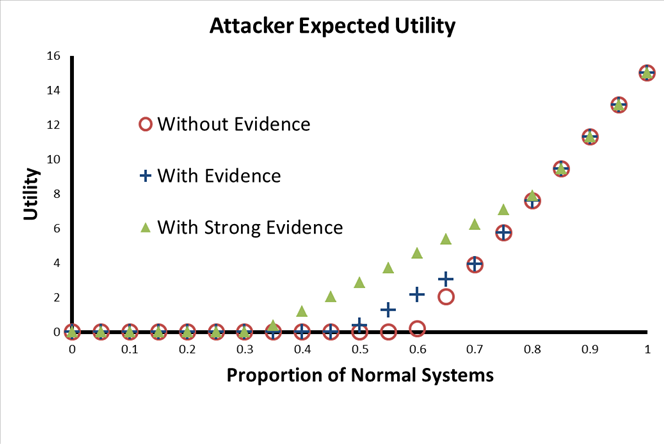}
\par\end{centering}

\protect\caption{\label{fig:uSAttacker1}Expected utility for the attacker in games
of $\mathcal{G}_{honey}$ and $\mathcal{G}_{honey}^{evidence}$ as
a function of the fraction $p\left(0\right)$ of normal systems in
the network.}

\begin{centering}
\includegraphics[width=0.6\columnwidth]{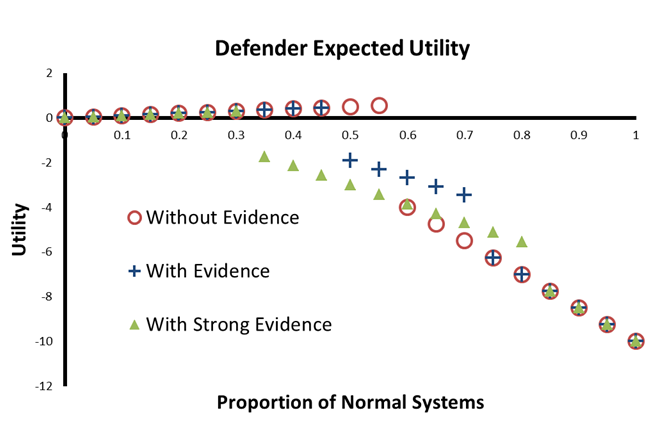}
\par\end{centering}

\protect\caption{\label{fig:uDefender}Expected utility for the defender in games of
$\mathcal{G}_{honey}$ and $\mathcal{G}_{honey}^{evidence}$ as a
function of the fraction $p\left(0\right)$ of normal systems in the
network.}

\end{figure}

First, consider the case in which the attacker does not have the ability
to detect honeypots, \emph{i.e. }$\mathcal{G}_{honey}$. The parameters
which determine the attacker and defender utilities are set according
to Table \ref{tab:Sample-parameters-which}. The attacker's utility
as a function of the fraction of normal systems in the network is
given by the red (circular) data points in Fig. \ref{fig:uSAttacker1}.
We can distinguish two parameter regions. When the proportion of honeypots
in the network is greater than approximately $40\%$, (\emph{i.e.}
$p\left(0\right)<60\%$), the attacker is completely deterred. Because
of the high likelihood that he will encounter a honeypot if he attacks,
he chooses to withdraw from all systems. As the proportion of normal
systems increases after $p\left(0\right)>60\%$, he switches to attacking
all systems. He attacks regardless of the sender's signal, because
in the pooling equilibrium, his signal does not convey any information
about the type to the receiver. In this domain, as the proportion
of normal systems increases, the expected utility of the attacker
increases. 

For this case in which the attacker cannot detect honeypots, the defender's
expected utility as a function of $p\left(0\right)$ is given by the
red (circular) data points in Fig. \ref{fig:uDefender}. We have noted
that, in the domain $p\left(0\right)<60\%$, the attacker always withdraws.
In this domain, it is actually beneficial for the defender to have
as close as possible to the transition density of $60\%$ normal systems,
because he gains more utility from normal systems that are not attacked
than from honeypots that are not attacked. But if the defender increases
the proportion of normal systems beyond $60\%$, he incurs a sudden
drop in utility, because the attacker switches form never attacking
to always attacking. Thus, the if the defender has the capability
to design his network with any number of honeypots, he faces an optimization
in which he wants to have as close as possible to $40\%$ of systems
be normal %
\footnote{At this limit, the defender's utility has a jump, but the attacker's
does not. It costs very little extra for the attacker to switch to
always attacking as $p\left(0\right)$ approaches the transition density.
Therefore, the defender should be wary of an ``malicious'' attacker
who might decide to incur a small extra utility cost in order to inflict
a large utility cost on the defender. A more complete analysis of
this idea could be pursued with multiple types of attackers. %
}. 

\subsection{Attacker Capable of Honeypot Detection}

Consider now how the network defense is affected if the attacker gains
some ability to detect deception. This game takes the form of $\mathcal{G}_{honey}^{evidence}$.
Recall that, in this form, a chance move has been added after the
sender's action. The chance move determines whether the receiver observes
evidence that the sender is being deceptive. For Fig. \ref{fig:uSAttacker1}
and Fig. \ref{fig:uDefender}, we have set the detection rates at
$\epsilon=0.8$ and $\delta=0.5$. These fall within the attacker
capability range $\mathcal{R}_{intermediate}$. Observing evidence
does not guarantee deception; neither does a lack of evidence guarantee
truth-revelation.

In the blue (cross) data points in Fig. \ref{fig:uSAttacker1}, we
see that, at the extremes of $p\left(0\right)$, the utility of the
attacker is unaffected by the ability to detect deception according
to probabilities $\epsilon$ and $\delta$. The low ranges of $p\left(0\right)$,
as described in table \ref{tab:Equilibria-for-Selected}, correspond
to the \emph{honeypot-saturated} region. In this region, honeypots
predominate to such an extent that the attacker is completely deterred
from attacking. Note that, compared to the data points for the case
without deception detection, the minimum proportion of honeypots which
incentives the attacker to uniformly withdraw has increased. Thus,
for instance, a $p\left(0\right)$ of approximately $0.50$ incentivizes
an attacker without deception detection capabilities to withdraw from
all systems, but does not incentivize an attacker with deception detection
capabilities to withdraw. At $p\left(0\right)=0.50$, the advent of
honeypot-detection abilities causes the defender's utility to drop
from $0.5$ to approximately $-2$. At the other end of the $p\left(0\right)$
axis, we see that a high-enough $p\left(0\right)$ causes the utilities
to again be unaffected by the ability to detect deception. This is
because the proportion of normal systems is so high that the receiver's
best strategy is to attack constantly (regardless of whether he observes
evidence for deception).

In the middle (non-saturated)\emph{ }region of $p\left(0\right)$,
the attacker's strategy is no longer to solely attack or solely withdraw.
This causes the ``cutting the corner'' behavior of the attacker's
utility in Fig. \ref{fig:uSAttacker1}. This conditional strategy
also induces the middle region for the defender's utility in Fig.
\ref{fig:uDefender}. Intuitively, we might expect that the attacker's
ability to detect deception could only decrease the defender's utility.
But the middle (\emph{non-saturated}) range of $p\left(0\right)$
shows that this is not the case. Indeed from approximately $p\left(0\right)=0.6$
to $p\left(0\right)=0.7$, the defender actually benefits from the
attacker's ability to detect deception! The attacker, himself, always
benefits from the ability to detect deception. Thus, there is an interesting
region of $p\left(0\right)$ for which the ability of the attacker
to detect deception results in a mutual benefit.

Finally, we can examine the effect of evidence as it becomes more
powerful in the green (triangle) points in Fig. \ref{fig:uSAttacker1}
and Fig. \ref{fig:uDefender}. These equilibria were obtained for
$\epsilon=0.9$ and $\delta=0.3$. This more powerful detection capability
broadens the middle parameter domain in which the attacker bases his
strategy partly upon evidence. Indeed, in the omnipotent detector
case, the plots for both attacker and defender consist of straight
lines from their utilities at $p\left(0\right)=0$ to their utilities
at $p\left(0\right)=1$. Because the attacker with omnipotent detector
is able to discern the type of the system completely, his utility
grows in proportion with the proportion of normal systems, which he
uniformly attacks. He withdraws uniformly from honeypots.

\section{Related Work\label{sec:Related-Work}}

Deception has become a critical research area, and several works have
studied problems similar to ours. Alcan et al. \cite{key-27} discuss
how to combine sensing technologies within a network with game theory
in order to design intrusion detection systems. They study two models.
The first is a cooperative game, in which the contribution of different
sensors towards detecting an intrusion determines the coalitions of
sensors whose threat values will be used in computing the threat level.
In the second model, they include the attacker, who determines which
subsystems to attack. This model is a dynamic (imperfect) information
game, meaning that as moves place the game in various information
sets, players learn about the history of moves. Unlike our model,
it is a complete information game, meaning that both players know
the utility functions of the other player. 

Farhang et al. study a multiple-period, information-asymmetric attacker-defender
game involving deception \cite{key-5}. In their model, the sender
type - benign or malicious - is known only with an initial probability
to the receiver, and that probability is updated in a Bayesian manner
during the course of multiple interactions. In \cite{key-25}, Zhuang
et al. study deception in multiple-period signaling games, but their
paper also involves resource-allocation. The paper has interesting
insights into the advantage to a defender of maintaining secrecy.
Similar to our work, they consider an example of defensive use of
deception. In both \cite{key-5} and \cite{key-25}, however, players
update beliefs only through repeated interactions, whereas one of
the players in our model incorporates a mechanism for deception detection. 

We have drawn most extensively from the work of Carroll and Grosu
\cite{key-17}, who study the strategic use of honeypots for network
defense in a signaling game. The parameters of our attacker and defender
utilities come from \cite{key-17}, and the basic structure of our
signaling game is adapted from that work. In \cite{key-17}, the type
of a particular system is chosen randomly from the distribution of
normal systems and honeypots. Then the sender chooses how to describe
the system (as a normal system or as a honeypot), which may be truthful
or deceptive. For the receiver's move, he may choose to attack, to
withdraw, or to condition his attack on testing the system. In this
way, honeypot detection is included in the model. Honeypot detection
adds a cost to the attacker regardless of whether the system being
tested is a normal system or a honeypot, but mitigates the cost of
an attack being observed in the case that the system is a honeypot.
In our paper, we enrich the representation of honeypot testing by
making its effect on utility endogenous. We model the outcome of this
testing as an additional move by nature after the sender's move. This
models detection as technique which may not always succeed, and to
which both the sender and receiver can adapt their equilibrium strategies.

\section{Discussion\label{sec:Discussion}}

In this paper, we have investigated the ways in which the outcomes
of a strategic, deceptive interaction are affected by the advent of
deception-detecting technology. We have studied this problem using
a version of a signaling game in which deception may be detected with
some probability. We have modeled the detection of deception as a
chance move that occurs after the sender selects a message based on
the type that he observes. For the cases in which evidence is trivial
or omnipotent, we have given the analytical equilibrium outcome, and
for cases in which evidence has partial power, we have presented numerical
results. Throughout the paper, we have used the example of honeypot
implementation in network defense. In this context, the technology
of detecting honeypots has played the role of a malicious use of anti-deception.
This has served as a general example to show how equilibrium utilities
and strategies can change in games involving deception when the agent
being deceived gains some detection ability.

Our first contribution is the model we have presented for signaling
games with deception detection. We also show how special cases of
this model cause the game to degenerate into a traditional signaling
game or into a complete information game. Our model is quite general,
and could easily be applied to strategic interactions in interpersonal
deception such as border control, international negotiation, advertising
and sales, and suspect interviewing. Our second contribution is the
numerical demonstration showing that pure-strategy equilibria are
not supported under this model when the distribution of types is in
a middle range but are supported when the distribution is close to
either extreme. Finally, we show that it is possible that the ability
of a receiver to detect deception could actually increase the utility
of a possibly-deceptive sender. These results have concrete implications
for network defense through honeypot deployment. More importantly,
they are also general enough to apply to the large and critical body
of strategic interactions that involve deception.

\end{document}